\documentclass[prb,showpacs,superscriptaddress,footinbib]{revtex4}

\usepackage{epsfig}
\usepackage{dcolumn}
\usepackage{bm}
\usepackage{amsmath}
\usepackage{amsfonts}
\usepackage{amssymb}
\usepackage{graphicx}%
\newcommand{\bP}{\bar{\Psi}}

\newcommand{\slA}{\raise.15ex\hbox{$/$}\kern-.57em\hbox{$A$}}
\newcommand{\slB}{\raise.15ex\hbox{$/$}\kern-.57em\hbox{$B$}}
\newcommand{\bA}{\bar{\alpha}}
\newcommand{\bB}{\bar{\beta}}
\newcommand{\beq}{\begin{equation}}
\newcommand{\eeq}{\end{equation}}
\newcommand{\beqn}{\begin{eqnarray}}
\newcommand{\eeqn}{\end{eqnarray}}
\newcommand{\slp}{\raise.15ex\hbox{$/$}\kern-.57em\hbox{$\partial$}}
\newcommand{\pslash}{\raise.15ex\hbox{$/$}\kern-.57em\hbox{$p$}}
\newcommand{\bX}{\bar{\chi}}

\begin{document}

\title{Spin correlation function in 2D statistical mechanics models with inhomogeneous line defects}

\author{Carlos Na\'on}
\author{Marta Trobo}
\affiliation{Instituto de F\'\i sica La Plata, CCT La Plata, CONICET
and Departamento de F\'\i sica, Facultad de Ciencias Exactas,
Universidad Nacional de La Plata, CC 67, 1900 La Plata, Argentina}

\date{December, 2010}

\begin{abstract}
We consider the critical spin-spin correlation function of the 2D
Ising model with a line defect which strength is an arbitrary
function of position. By using path-integral techniques in the
continuum description of this model in terms of fermion fields, we
obtain an analytical expression for the correlator as functional of
the position dependent coupling. Thus, our result provides one of
the few analytical examples that allows to illustrate the transit of
a magnetic system from scaling to non-scaling behavior in a critical
regime. We also show that the non-scaling behavior obtained for the
spin correlator along a non-uniformly altered line of an Ising model
remains unchanged in the Ashkin-Teller model.
\end{abstract}
\maketitle

\section{Introduction}
Two dimensional statistical mechanics systems play a central role in
our present understanding of phase transitions and critical
phenomena. Outstanding members of this family of theories are the
Ising model, the Ashkin-Teller \cite{AT} and the eight-vertex
\cite{Bax} models. These models are useful to shed light on a
variety of phenomena, in both classical and quantum physics, ranging
from biological applications \cite{DNA} to the theory of cuprate
superconductors \cite{cuprate}. Moreover, important advances in
material science, accomplished over the last decades, have developed
the ability to grow and experimentally explore ultrathin
ferromagnetic films \cite{experiments}, giving the opportunity to
test some of the theoretical predictions. One of the fundamental
questions concerning these essentially 2D materials is the role of
defects and impurities in the critical properties of magnetic
systems. Apart from academic interest, a detailed knowledge on the
influence of defects on physical properties is always useful on
general grounds, since all real materials are, to some extent,
defected. In some cases of applied interest, such as
ultrahigh-density magnetic recording media, it has been shown that
linear defects can be used to efficiently control domain wall
pinning, thus stabilizing the large area domain structure of
ultrathin films \cite{defects-recording}. Linear charge defects may
also appear in graphene grown by chemical vapor deposition on Ni
surface \cite{graphene}.

On the theoretical side, very little is known exactly about the
behavior of planar systems in the presence of line defects
\cite{Igloi-Peschel-Turban}. For the simple square Ising lattice
with an altered row (Bariev's model \cite{Bariev}) it has been shown
that the scaling index of the magnetization varies continuously with
the defect strength \cite{Bariev, McCoy-Perk}, whereas the critical
exponent of the energy density at the defect line remains unchanged
\cite{Brown, Ko-Yang-Perk, Burkhardt-Choi}. Taking this model as
working bench, much insight was obtained about the origin of
nonuniversal critical behavior \cite{Burkhardt-Eisenriegler}. More
recently, by using path-integrals within the continuous formulation
of Ashkin-Teller and Baxter models, it was shown that the magnetic
exponent depends on the strength of the defect in exactly the same
way as in Bariev's model \cite{naon09}.

From another perspective, due to the well-known connection between
the classical 2D Ising model and a quantum field theory of Dirac
fermions in $1+1$ dimensions, the study of defects as perturbations
of conformal field theories has led to very important results in the
area of integrable quantum field theories \cite{Mussardo, Goshal,
Konik}. This line of research was later focused on the problem of
conductance in quantum wires \cite{Castro}. The analysis of more
mathematical aspects concerning the role of impurities and defects
in the renormalization group flows of conformal models is currently
under intense investigation \cite{Fendley, Kormos}. Very recently,
the entanglement between two pieces of a quantum chain was analyzed
by exploiting the connection with an Ising model with a defect line
\cite{Eisler-Peschel}.

All these advances were achieved for the case of homogeneous
defects, i.e. when the defect strength is constant along the altered
line. The case of non uniform couplings has been analyzed in the
context of extended defects at surfaces \cite{extended, extended1,
extended2} and in the bulk \cite{extendedBariev, extendedIgloi},
displaying a rich variety of behaviors in the local critical
properties.

In this work we consider a narrow inhomogeneous defect and study the
spin-spin correlator on the altered line. In other words, we analyze
the extension of Bariev's model to the case in which the strength of
the line defect is a function of the position on the column with
modified couplings. Then, our result for the critical spin-spin
correlator is a generalization of the result first obtained in
Ref.[\onlinecite{McCoy-Perk}] for a uniform line defect. By using a
path-integral approach in the continuum limit, we have obtained a
formula that gives the spin-spin correlation function as a
functional of an arbitrary defect distribution. This allows to
explore the effect of different types of specific alterations in a
straightforward way. We have also shown that the results remain
valid for the Ashkin-Teller model, i.e. we found that in these
altered systems the non-scaling behavior of magnetic correlations on
the inhomogeneous defect coincides with the one obtained in the
Ising case.

The paper is organized as follows. In Section II we explain our
computational procedure for the well-known defect-free Ising model.
In Section III we show how to extend the method when a line of
altered couplings is included in the system. We emphasize how the
case of inhomogeneous defect strength can be naturally considered
with our technique. In Section IV we illustrate the use of our
result showing the predictions for two specific defect functions. In
Section V we extend the procedure to the more complex Ashkin-Teller
and Baxter models.  Finally in Section VI we summarize our findings
and present our conclusions.

\section{The method: Defect-free case}

For completeness and illustrative purposes, we start by describing
the computational procedure for the homogeneous defect-free case.
The Hamiltonian of the original lattice model is given by
\begin{equation}
{\cal H} = -\sum_{<ij>}J_2 \, \sigma_i \sigma_j
\end{equation}
where $<ij>$ means that the sum runs over nearest neighbors of a
square lattice ($\sigma=\pm 1$).

As shown in Ref. [\onlinecite{BI}] the scaling regime of the 2D IM
can be described in the continuum limit in terms of a model of
Majorana fermions with Lagrangian density:
\begin{equation}\label{lagrangian}
{\cal L}[\alpha] = \bA i\slp \alpha
\end{equation}
where $\alpha$ represents a Majorana spinor with components
$\alpha_{1,2}$. Let us recall that this components are connected to
fermion annihilation and creation operators $c_r$ and
$c^{\dagger}_r$ attached to site $r$ ($c_r =
\frac{e^{-i\pi/4}}{\sqrt{2}}(\alpha_{1}(r)+i\alpha_{2}(r)$). It is
also useful for later convenience to define the energy-density as
$\epsilon_{\alpha}=\alpha_{1} \alpha_{2}$. The symbol $\slp$ stands
for $\gamma_{\nu}\partial_{\nu}$, with $\gamma_{\nu}$ the usual
Euclidean Dirac matrices ($\nu=0,1$ associated to space directions).

Similar manipulations, based on the Jordan-Wigner transformation
\cite{SML}, allow to write the on-line spin-spin correlation
function in the form \cite{BI}
\begin{equation}\label{correlation}
<\sigma(0)\sigma(R)>_{Ising} = <\exp\;({\pi \int_{0}^{R}dx\,
\epsilon_{\alpha}(x)})>
\end{equation}
where the vacuum expectation value is an anticommuting path-integral
to be evaluated with the continuum action $S = \int d^2x\,{\cal L}$,
with an integration measure $\cal{D}\alpha$. The explicit
computation of (\ref{correlation}) can be performed either in terms
of the Majorana $\alpha$-fields or in terms of Dirac fermions
\cite{ZI} built through the doubling technique \cite{Ferrell},
yielding the well-known result for the Ising correlator. We start by
squaring (\ref{correlation}):
\begin{equation}\label{squared}
<\sigma(0)\sigma(R)>_{Ising}^2 = <\exp\;\big({ \pi \int_{0}^{R}dx\,
(\epsilon_{\alpha}(x)+\epsilon_{\alpha'}(x))}\big)>
\end{equation}
where the vacuum expectation value must now be computed with respect
to an Euclidean action with Lagrangian density $\tilde {\cal
L}[\alpha,\alpha']= {\cal L}[\alpha]+ {\cal L}[\alpha']$, $\alpha'$
being the replicated fermion fields. Following Ref.[\onlinecite{ZI}]
we can build Dirac fermions $\Psi$ as
\begin{equation}\label{diracfermion}
\Psi= \alpha + i \alpha'.
\end{equation}

In terms of these new fields we can write the Lagrangian density
$\tilde {\cal L}[\alpha,\alpha']$ in the form
\beqn\label{newLagrangian}&\tilde {\cal L}[\Psi]&=\bP i\slp \Psi ,
\eeqn where $\gamma_5 =i\gamma_0 \gamma_1$. On the other hand
equation (\ref{squared}) can be expressed as
\begin{equation}\label{final-squared}
<\sigma(0)\sigma(R)>_{Ising}^2 = <\exp\;\Big({\pi \int
d^2x\,\bP\,\slA\,\Psi}\Big)>,
\end{equation}
where now the path integral integration measure in the right hand
side is expressed in terms of the fields $\Psi$, and $A_{\nu}$ is an
auxiliary vector field with components: \beqn \label{Amu1}
A_{0}(x_{0},x_{1}) = -\delta(x_{0}) \theta(x_{1}) \theta(R-x_{1}),
\,\,A_{1}(x_{0},x_{1}) = 0 . \eeqn Gathering the above results we
can write:
\begin{equation}\label{partitions}
<\sigma(0)\sigma(R)>_{Ising}^2 = \frac{Z[g=\pi]}{Z[g=0]},
\end{equation}
where \begin{equation} Z[g]=\int{\cal D}\bP{\cal
D}\Psi\,\exp\big({-\int d^{2}x\big(\tilde {\cal L}[\Psi]+ g \bP
\slA\,\Psi \big)}\big).\end{equation}

The continuum limit of the squared two-point spin correlation
function is {\em exactly} expressed in terms of the vacuum to vacuum
functional of a quantum field theory describing a Dirac fermion
interacting with a classical background $A_{\nu}$. Now we make the
following change of path-integral variables in the numerator of
equation (\ref{partitions}), with chiral and gauge parameters $\Phi$
and $\eta$, respectively: \beq \label{changeofvar1} \Psi =
e^{-\pi(\gamma_{5}\Phi-i\eta)}\;\zeta,\,\, \bP = {\bar{\zeta}}\;
e^{-\pi(\gamma_{5}\Phi+i\eta)}. \eeq

The integration measures ${\cal D}\Psi$ and ${\cal D}\zeta$ are
related through the so called Fujikawa Jacobian $J$, ${\cal
D}\bP\,{\cal D}\Psi= J[\Phi,\eta]\,{\cal D}{\bar{\zeta}}\,{\cal
D}\zeta$. If the parameters of the transformation are related to the
previously introduced vector field $A_{\nu}$ in the form \beq
\label{Amu}A_{\nu} = \epsilon_{\nu\rho}\partial_{\rho}\Phi +
\partial_{\nu}\eta \eeq
one easily gets $Z[g=\pi]=J\,Z[g=0]$, which leads to
\begin{equation}\label{jacob}
<\sigma(0)\sigma(R)>_{Ising}^2 = J(R),
\end{equation}

As explained in Ref. [\onlinecite{Nao}], the Jacobian $J(R)$ must be
computed with a gauge-invariant regularization prescription in order
to avoid an unphysical linear divergence. Following this procedure
one finds that $J$ depends on the $\Phi$-field only as
\begin{equation}\label{Jacobian}
J(R)=\exp{-\pi/2\int
d^2x\,\partial_{\nu}\Phi(x,R)\,\partial^{\nu}\Phi(x,R)}.
\end{equation}

The explicit form of $\Phi(x,R)$ is determined by combining Eqs.
(\ref{Amu1}) and (\ref{Amu}) which gives the following partial
differential equation for $\Phi$:
\begin{equation}
\Box\Phi(x_0,x_1,R)=
-\delta(x_0)\,\frac{d}{dx_1}[\theta(x_1)\theta(R-x_1)]
\end{equation}
where $\Box=\partial_0^2 +\partial_1^2$. The solution of this
equation is easily obtained by using the Green's function of the
D'Alembertian: $G_0(z_0,z_1)=\frac{1}{4\pi}\ln(z_0^2 + z_1^2 +
a^2)$, with $a$ an ultraviolet cutoff related to the original
lattice spacing. Replacing in (\ref{Jacobian}) and considering the
limit $R>>a$ we find the well-known result
 $<\sigma(0)\sigma(R)>_{Ising}\simeq
(a/R)^{1/4}$.

\section{Inhomogeneous line defect}

Now we include a line defect in the original Ising lattice. To be
specific we consider the so called chain defect (here we employ the
terminology of Ref. [\onlinecite{Igloi-Peschel-Turban}], which
corresponds to Bariev's second type defect, in which bonds along the
same column are replaced: $J_2 \rightarrow J_2'$). In previous
studies the altered coupling $J_2'$ was taken as a constant. From
now on we allow $J_2'$ to vary from site to site, i.e. we make $J_2'
\rightarrow J_2'(x_1)$.

We will study the two-spin correlation function in the column of
altered bonds ($x_0=0$) \cite{McCoy-Perk}. It is known that the
continuous version of the classical model is modified, due to the
defect, by the addition in equation (\ref{lagrangian}) of a term
$2\pi\mu(x_1)\,\delta(x_0)\,\epsilon_{\alpha}(x)$, with
$\mu=J_2'(x_1)-J_2$ (see for instance
[\onlinecite{Burkhardt-Choi}]). By carefully examining the fermionic
representation of $\sigma$-spin operators on the lattice, following
the lines of Ref. [\onlinecite{ZI}], one also finds that in the
continuum limit each spin operator on the defect line picks up a
similar $\mu$-dependent factor, in such a way that the correlator
for the defective model is given by a simple modification of
equation (\ref{correlation}):
\begin{equation}
<\sigma(0)\sigma(R)>_{inhom} = <\exp\big(\pi \int
dx_{1}\,(1+4\mu(x_1))\,\epsilon_{\alpha}(x_{1})\big)>_{\mu}.
\end{equation}

It is evident that the squared correlator can be written again as in
equation (\ref{final-squared}). The presence of the inhomogeneous
defect manifests in the form of the $A_{\nu}$-field which is now
rescaled by a factor $(1+4\mu(x_1))$. Thus equation (\ref{Amu1})
becomes
\begin{equation}\label{AInhom}
A_{0}(x_{0},x_{1}) = -(1+4\mu(x_1))\delta(x_{0}) \theta(x_{1})
\theta(R-x_{1})
\end{equation}

\begin{equation}
A_{1}(x_{0},x_{1}) = 0.
\end{equation}

The implementation of the change of variables given by
(\ref{changeofvar1}) and (\ref{Amu}) leads to the generalization of
(\ref{jacob}):
\begin{equation}
<\sigma(0)\sigma(R)>_{inhom}^2 = J_{inhom}(R).
\end{equation}
Formally $J_{inhom}(R)$ is still given by (\ref{Jacobian}), but the
effects coming from the nonuniformity of the defect strength enters
the game through the function $\Phi(x_0,x_1,R)$, which now obeys a
non trivial differential equation depending on $\mu(x_1)$:
\begin{equation}
\Box\Phi(x_0,x_1,R)=
-\delta(x_0)\,\frac{d}{dx_1}[(1+4\mu(x_1))\,\theta(x_1)\,\theta(R-x_1)].
\end{equation}
The formal solution of this equation is

\beqn &\Phi(x_0,x_1,R)=\frac{1}{4\pi}\ln{\frac{x_0^2 + a^2 +
(x_1-R)^2}{x_0^2 + a^2 + x_1^2}}+ \nonumber\\&+\frac{1}{\pi}\int_0^R
dx_1'\,\mu(x_1')\,\frac{d}{dx_1'}\ln{[x_0^2 + (x_1-x_1')^2 + a^2]}.
\eeqn

Replacing in the corresponding expression for $J_{inhom}(R)$ we
obtain

\beqn \label{sigma-inhom}&<\sigma(0)\sigma(R)>_{inhom}=
(\frac{a^2}{a^2+R^2})^{\frac{1}{8}+\frac{\mu(0)+\mu(R)}{4}}\,
e^{F(R)},\eeqn

where

\beqn \label{F}&F(R)=\frac{1}{4}\int_0^R
dx\,\mu(x)\,\frac{d}{dx}\big[ \ln{\frac{(a^2 +
(x-R)^2)^{(1+4\mu(R))}}{(a^2 + x^2)^{(1+4\mu(0))}}}\big]
-\nonumber\\&- \frac{1}{4}\int_0^R\int_0^R
dx\,dy\,(1+4\mu(x))\,\frac{d}{dy}\mu(y)\,\frac{d}{dx}\big[\ln{(a^2 +
(x-y)^2)}\big].\eeqn In the above integrals we have dropped the
subindex 1 in the integration variables, in order to simplify the
notation ($x_1 \rightarrow x$ and $y_1 \rightarrow y$). It is easy
to check that in the special case $\mu(x)\rightarrow\mu=constant$,
one obtains
\begin{equation}\label{correl-defective}
<\sigma(0)\sigma(R)>_{\mu} \simeq (\frac{a}{R})^{2\Delta_{\sigma}},
\end{equation}
with $\Delta_{\sigma}=\frac{1}{8}(1+4\mu)^2$, which is the
well-known result first obtained by McCoy and Perk
\cite{McCoy-Perk}.

Formulae (\ref{sigma-inhom}) and (\ref{F}) constitute the main
formal result of this paper. They give the critical spin-spin
correlation on the altered line of an Ising model, as a functional
of an arbitrarily varying defect strength. In the next section we
will show some specific predictions for definite defect
distributions.

\section{Application to some specific defects}

\begin{figure}[h]\begin{center}
\includegraphics[scale=1]{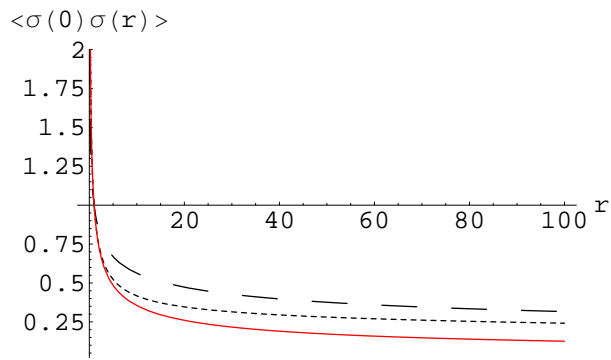}
\caption{\label{figu1}Spin-spin correlation as a function of $r$ for
constant (solid line) and variable (pointed line) defect strength.
We set $\beta = 10$. The dashed line indicates the defect-free
universal behavior $r^{-1/4}$.}
\end{center}\end{figure}

Let us now consider some specific defect-functions for which $F(R)$
can be analytically evaluated. We start with the following defect
distribution
\begin{equation}\label{function1}
\mu(x)= \mu_0 \frac{1}{(1 + \mid x \mid /b)}.
\end{equation}
where $b$ is a characteristic length scale. This function is similar
to the one considered by Bariev in his study of horizontal large
scale inhomogeneities \cite{extendedBariev}. Passing to
dimensionless variables $r=R/a$ and $\beta=b/a$, and considering
weak defect strengths ($\mu_0\ll1$), for $R\gg a$ and $b\gg a$ we
obtain

\beqn & <\sigma(0)\sigma(r)>_{inhom}=
\big(\frac{1}{r}\big)^{\frac{1}{4}+\frac{\mu_0(2\beta +
r)}{(\beta+r)}}\, \big(\frac{\beta}{\beta+r}\big)^{\frac{-\mu_0
r}{(\beta+r)}}\, \exp^{\big(\frac{\mu_0  r (2\beta
+r)}{\beta(\beta+r)^2} \arctan(r)\big)} .\eeqn We see that the
magnetic correlation exhibits non-scaling behavior, as expected
for a local inhomogeneity. This result is in qualitative agreement
with the analysis of Ref.[\onlinecite{extendedBariev}]. However we
should stress that we are considering a different situation here.
Indeed, the present case corresponds to a standard 2D Ising model
in which just one column ($x_0=0$) is altered in a non uniform
way, whereas in Ref.[\onlinecite{extendedBariev}] the couplings
along columns are kept constant, while the couplings along
\textit{all} rows are modified in a non uniform fashion. In Figure
\ref{figu1} we compare the decays of correlations for constant
defect (solid line), non constant defect with decay law
(\ref{function1}) (pointed line) and the universal defect-free
behavior (dashed line). In agreement with physical intuition the
correlation decays monotonically with distance, in an intermediate
way, faster than the defect-free case and slower than the case in
which the defect strength is constant. Another expected feature,
well reproduced by our solution, concerns the behavior with
$\beta=b/a$: for increasing $\beta$ the non-scaling decay becomes
faster, being undistinguishable from the uniform case for large
enough $\beta$.

\vspace*{1cm}

\vspace*{1cm}

Let us now study a different function $\mu(x)$, which represents a
non-monotonic alteration of the line $x_0=0$. For simplicity we
consider a sequence of $N$ slabs of heights $\mu_{0 i}$
($i=1,...N$). Each slab starts at $x=a c_i$ and ends at $x=a d_i$:

\begin{equation}
\mu(x)= \sum_{i=1}^N \mu_{0i}\,\theta(x -a c_{i})\,\theta(a d_{i}-
x),
\end{equation}
where $\theta(x)$ is Heaviside's function. Evaluating $F(R)$ and
replacing in (\ref{sigma-inhom}), in the weak coupling regime
($\mu_{0i}\ll1$) and for $r, c_i, d_i \gg1$ we obtain

\beqn \label{multitetas}<\sigma(0)\sigma(r)>_{inhom} =
\big(\frac{1}{ r}\big)^{1/4}\prod_{i=1}^N \big[\big(\frac{1}{
r}\big)^2 \big(\frac{c_{i}^2}{(c_{i} - r)^2 + 1}\big)\big]^{\mu_{0i}
\theta(d_{i}- r) \theta(r-c_{i})/2}\,
\big[\big(\frac{c_{i}}{d_{i}}\big)^2\big(\frac{(d_{i}-r)^2 +
1}{(c_{i}-r)^2+1}\big)\big]^{\mu_{0i}\theta(r-d_{i})/2} \eeqn

\begin{figure}[h]\begin{center}
\includegraphics{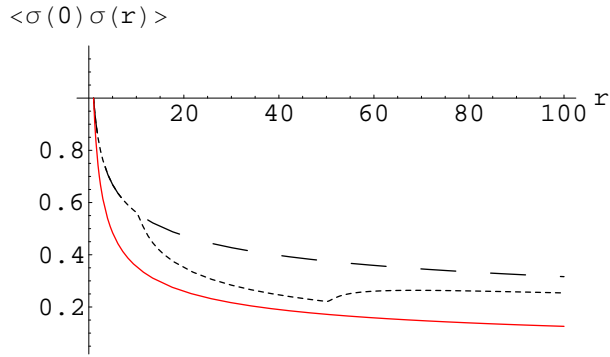}
\caption{\label{figu2}Spin-spin correlation as a function of $r$ for
a line defect given by a slab starting at $x/a=10$ and ending at
$x/a=50$, for $\mu_0 = 0.1$ (pointed line). The dashed line
indicates the defect-free universal behavior $r^{-1/4}$. The solid
line corresponds to a uniform defect with $\mu_0 = 0.1$.}
\end{center}\end{figure}

\begin{figure}[h]\begin{center}
\includegraphics{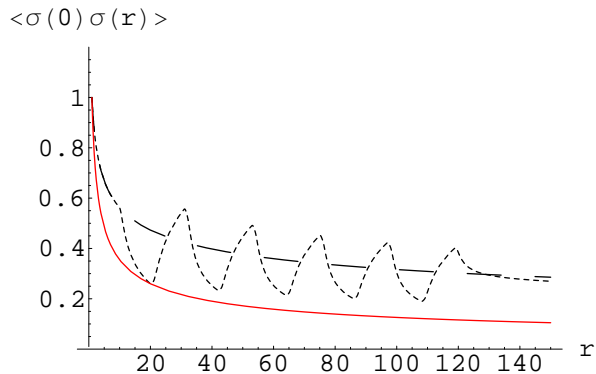}
\caption{\label{figu3}Spin-spin correlation as a function of $r$ for
a line defect given by an oscillatory function (a sequence of 5
slabs and wells with equal heights (depths) and widths
($\mid\mu_0\mid = 0.1$),starting at $x/a=10$ and ending at
$x/a=119$) (pointed line). The dashed line indicates the defect-free
universal behavior $r^{-1/4}$. The solid line corresponds to a
uniform defect with $\mu_0 = 0.1$.}
\end{center}\end{figure}

In Figure \ref{figu2} we display the result given by the above
formula for the simplest case: one slab or "barrier" starting at
$x/a=c=10$ and ending at $x/a=d=50$. For $r<c$ the critical two-spin
correlation coincides with the standard, non defected correlation.
In the presence of the defect, for $c<r<d$, it exhibits a faster
decay. The correlation reaches a local minimum at $r=d$, and then it
starts growing, approaching again the universal behavior
corresponding to the magnetic critical index $1/8$, asymptotically.
In Figure \ref{figu3}, taking into account that (\ref{multitetas})
is valid for both positive and negative values of $\mu_0$, we show
the critical correlation for a defect which is oscillatory along
certain portion of the line $x_0=0$, a sequence of five alternated
slabs ($\mu_0=0.1$) and wells ($\mu_0=-0.1$). As before, the
spin-spin function coincides with the non defected one, for small
distances ($r<c_1$). For $c_1<r<d_5$ there is an oscillatory
behavior around the universal curve $r^{-1/4}$. For large distances
the correlation tends to the universal decay.

\vspace*{1cm}

\section{Extension to the Ashkin-Teller model}
In this Section we show how to extend the treatment of inhomogeneous
linear impurities to the Ashkin-Teller system consisting of two
Ising subsystems with spin variables $\sigma_i$ and $\tau_i$ coupled
by a quartic interaction \cite{AT} \cite{Bax}. The corresponding
lattice Hamiltonian reads
\begin{equation}
{\cal H} = -\sum_{<ij>}[J_2 \, (\sigma_i \sigma_j + \tau_i \tau_j) +
J_4 \,\sigma_i \sigma_j \tau_i \tau_j]
\end{equation}
where $<ij>$ means that the sum runs over nearest neighbors of a
square lattice ($\sigma, \tau,=\pm 1$). As it is well known, in the
vicinity of the critical point this model can be described in the
continuum limit in terms of two Majorana fermions interacting via
their energy-densities:
\begin{equation}\label{lagrangianAT}
{\cal L}[\alpha, \beta] = \bA i\slp \alpha + \bB i\slp \beta -
\lambda \, \epsilon_{\alpha}\,\epsilon_{\beta}
\end{equation}
where $\alpha$ and $\beta$ are the Majorana spinors with components
$\alpha_{1,2}$, $\beta_{1,2}$ respectively.
$\epsilon_{\alpha}=\alpha_{1} \alpha_{2}$ and
$\epsilon_{\beta}=\beta_{1} \beta_{2}$ are the corresponding
energy-densities. The coupling constant $\lambda$ is proportional to
$J_4/J_2$. Let us now include, as before, a linear defect affecting
one of the original Ising lattices, say the one with spins $\sigma$.
If this impurity is placed at column $x_0=0$, in the continuum limit
we have to add to ${\cal L}$ a term
$2\pi\mu(x_1)\,\delta(x_0)\,\epsilon_{\alpha}(x)$, with
$\mu=J_2'-J_2$. As shown in Ref. \onlinecite{naon09}, in order to
compute the spin-spin correlator on the altered line is still
possible to use the doubling technique depicted in Section II.
However, in spite of the formal analogy, the situation is much more
complex here. First of all, since we have two sets of spins, we have
to introduce two Dirac fields: $\Psi= \alpha + i \alpha'$ and $\chi=
\beta + i \beta'$. We then obtain
\begin{equation}\label{string}
<\sigma(0)\sigma(R)>_{AT}^2 = <\exp\Big({\pi \int
d^2x\,\,\bP\,\slA\,\Psi}\Big)>_{\mu}.
\end{equation}
Here the background field $A_{\nu}$ is given by (\ref{AInhom}) and
the vacuum expectation value must be computed with respect to an
Euclidean action with Lagrangian density $ {\cal L}[\Psi, \chi]$:

\beqn \tilde {\cal L}[\Psi, \chi] & = & \bP i\slp \Psi + \bX i\slp
\chi \nonumber \\ & - & \frac{\lambda}{8} [ \bX \gamma_5 \chi \bP
\gamma_5 \Psi + Im(\chi^T \gamma_1 \chi) Im(\Psi^T \gamma_1 \Psi)],
\eeqn where $\gamma_5 =i\gamma_0 \gamma_1$ and $\Psi^T, \chi^T$ are
the transposed spinors.

The implementation of the change of variables given by
(\ref{changeofvar1}) and (\ref{Amu}) leads to

\begin{equation}\label{correlator}
<\sigma(0)\sigma(R)>_{AT}^2 = <\sigma(0)\sigma(R)>_{inhom}^2
F(\lambda, R, \mu)
\end{equation}
where $<\sigma(0)\sigma(R)>_{inhom}$ is the defected Ising
correlator given in (\ref{sigma-inhom}) and
\begin{equation}
F(\lambda, R, \mu)= {\cal N}(\lambda) < \exp[ S_{\Phi}(\zeta,\chi) +
S_{\eta}(\zeta,\chi)]>_0
\end{equation}
where $< >_0$ means vacuum expectation value with respect to the
model of free $\chi$ and $\zeta$ fermions. ${\cal N}(\lambda)$ is a
normalization constant independent of R. Since the analysis of the
dependence of $F(\lambda,\mu,  R)$ on R is more easily done in
momentum space, we have Fourier transformed
$S_{\Phi}(\zeta,\chi,\mu)$ and $S_{\eta}(\zeta,\chi,\mu)$ in the
above equation:
\begin{equation}
S_{\Phi}(\zeta,\chi,\mu)= \frac{\lambda}{8}\int \prod_{j=1}^{4}
\frac{d^2p_j}{(2 \pi)^2}[ \bX(p_1) \gamma_5 \chi(p_2)
\bar{\zeta}(p_3)  \gamma_5 G(P,R,\mu)\zeta(p_4),
\end{equation}
with $G(P,R,\mu)$ being a diagonal 2x2 matrix given by
\begin{equation}
G(P,R,\mu) = \left(\begin{array}{cc}g_+(P,R,\mu) & 0 \\ 0 &
g_-(P,R,\mu)
\end{array}\right),
\end{equation}
where $g_{\pm}(P,R,\mu)= \pm \int d^2x e^{iPx} e^{\mp2 \pi
\Phi(x,\mu,R)}$ and $P= p_1+ p_2+p_3+p_4$. A similar expression is
obtained for $S_{\eta}$ with $G(P,R)$ replaced by
\begin{equation}
H(P,R,\mu) = \left(\begin{array}{cc}h(P,R,\mu) & 0 \\ 0 &
h(P,R,\mu)
\end{array}\right),
\end{equation}
with $h(P,R,\mu)= \int d^2x e^{iPx} e^{2 \pi \eta(x,R,\mu)}$. The
explicit functional forms of $\Phi(x,R,\mu)$ and $\eta(x,R,\mu)$ can
be determined following the same steps depicted in previous
Sections, yielding

\beqn &\Phi(x_0,x_1,R,\mu)=\frac{-1}{4\pi}\ln{\big(\frac{x_0^2 + a^2
+ (x_1-R)^2}{x_0^2 + a^2 + x_1^2}}\big)+
\nonumber\\&+\frac{2}{\pi}\int_0^R dx_1'\,\mu(x_1')\,\frac{(x_1 -
x_1')}{(x_0^2 + (x_1-x_1')^2 + a^2)} \eeqn and \beqn
&\eta(x_0,x_1,R,\mu)=\frac{x_0}{2\pi}\int_0^R dy \frac{(1 + 4
\mu(y))}{(x_0^2 + a^2 + (y - x_1)^2)}. \eeqn

Then,  $g(P,R,\mu)$ becomes
\begin{equation}
g_{\pm}(P,R,\mu)= \pm \int d^2x e^{iPx}\big(\frac{x_0^2 + a^2 +
(x_1-R)^2}{x_0^2 + a^2 + x_1^2}\big)^{\pm 1/2} e^{\mp 4\int_0^R dy
\mu(y) \frac{(x_1-y)}{(x_0^2 + a^2 + (y - x_1))^2} }
\end{equation}
and $h(P,R)$
\begin{equation}
h(P,R,\mu) = \int d^2x e^{iPx} e^{i x_0 /\sqrt{x_0^2 +a^2}\arctan
\big( \frac{R \sqrt{x_o^2+a^2}}{x_0^2+x_1^2+a^2-Rx_1}\big)} e^{i 4
x_0 \int_0^R dy \frac{\mu(y)}{(x_0^2 + a^2 + (y - x_1)^2)}}.
\end{equation}

Since any possible dependence on $R$ of the function $F(\lambda, R,
\mu)$ comes from $g_{\pm}(P,R,\mu)$ and $h(P,R,\mu)$, our problem is
reduced to the analysis of these integrals. Let us first introduce a
cutoff $L$, which can be interpreted as the size of the system, in
order to avoid infrared divergencies (the thermodynamic limit will
be recovered at the end of the computation by setting $L\rightarrow
\infty$). In terms of the dimensionless variable $u_\rho = x_\rho
/L$, ($\rho = 0,1$) we obtain \beqn g_{\pm}(P,R,\mu) & = &
\lim_{L\rightarrow \infty}  \pm L^2\int_{|u_\rho|<1} d^2u \nonumber
\\ &\times &e^{iPLu}\big( \frac{u_0^2 + a^2/L^2 +
(u_1-(R/L))^2}{u_0^2 + a^2/L^2 + u_1^2}\big)^{\pm 1/2} e^{\mp
\frac{4}{L}\int_0^R dy \mu(y)
\frac{(u_1-y/L)}{(u_0^2 + a^2/L^2 + (y/L - u_1))^2} }\nonumber\\
&=& \pm (2 \pi)^2 \delta^2(P) \eeqn and a similar result for
$h(P,R)$. Then, in the thermodynamic limit ($a\ll R \ll L$)
$F(\lambda, R, \mu)$ becomes independent of R and the critical
behavior coincides with the one of the 2D Ising model in presence of
an arbitrary inhomogeneous defect.

\section{Summary and conclusions}
We have considered the critical behavior of the two-spin correlation
function in the continuum, field-theory version of the 2D Ising
model with a line defect placed at the column $x_0=0$. In contrast
to previous studies, here we have taken into account possible
variations of the defect strength with the position on the line. Our
main result (Eqs. (\ref{sigma-inhom}) and (\ref{F})) provides an
analytical expression for the critical spin-spin correlation as a
functional of an arbitrary defect distribution. From this one can
explore the effect of different types of non uniform impurity
distributions on the magnetization. In particular our finding can be
used to analyze, within the critical regime, the transit from
scaling to non scaling behavior. As examples, in order to illustrate
the approach and check its validity, we have discussed two special
cases: a defect strength decaying monotonously with distance from a
given point, and a sequence of slabs. Finally, we extended the
analysis to a nonhomogeneous line defect placed at one column of an
Ashkin-Teller system, showing that the spin correlator on the
altered line decays, in the thermodynamic limit, in the same way as
in the Ising model.

\vspace{1cm}

\noindent {\bf Acknowledgement}\\
The authors are grateful to CONICET and UNLP (Argentina) for
financial support.

\end{document}